\documentstyle[epsfig]{elsart}

\def\be{\begin{equation}}
\def\ee{\end{equation}}
\def\bea{\begin{eqnarray}}
\def\eea{\end{eqnarray}}
\def\bk{{\mathbf k}}
\def\d{{\mbox{\rm d}}}

\begin{document}

\begin{frontmatter}
\title{\bf Bose-Einstein correlations\\
	for L\'evy stable source distributions }

\author[KFKI]{T. Cs\"org\H o,\thanksref{tamas}}
\author[KFKI]{S. Hegyi\thanksref{sandor}}
\author[Columbia]{ and W. A. Zajc\thanksref{bill}}
\address[KFKI]{KFKI RMKI, H-1525 Budapest 114, POB 49, Hungary}
\address[Columbia]{Dept. Physics, Columbia University, 538 W 120th St, NY 10027 New York}

\thanks[tamas]{Email: csorgo@sunserv.kfki.hu}
\thanks[sandor]{\phantom{Email:} hegyi@rmki.kfki.hu}
\thanks[bill]{\phantom{Email:} zajc@nevis.columbia.edu}


\begin{abstract}
The peak of the two-particle Bose-Einstein
correlation functions has a  very interesting structure. 
It is often believed to have a multivariate Gaussian form. 
We show here that for the class of stable distributions, 
characterized by the index of stability $0 < \alpha \le 2$, 
the peak has a stretched exponential shape. 
The Gaussian form corresponds then to the special 
case of $\alpha = 2$. We give examples for the Bose-Einstein
correlation functions for  univariate as well as  
multivariate stable distributions, and check the model against  
two-particle correlation data. 
\end{abstract}
\end{frontmatter}

\section{Introduction}
The structure of the peak of two-particle Bose-Einstein correlation functions
(BEC-s)
is a very interesting physically measurable quantity. It carries information
about the fraction of long-lived resonances~\cite{chalo}, hence signals possible
$U_A(1)$ symmetry restoration in high energy heavy ion reactions~\cite{bec-ua1},
it can be utilized to determine a generalized Hubble constant for the
transverse flow in high energy heavy ion physics~\cite{hubble}
and it is also sensitive to a possible partial coherence 
in the source~\cite{pcnhalo}, 
for other examples see the recent reviews of refs.~\cite{cs-rev,kittel-rev}.

Originally, the method of Bose-Einstein correlations
was invented by the radio astronomers 
R. Hanbury Brown and R. Q. Twiss (HBT), who applied it to determine 
the angular diameters of main sequence stars
~\cite{HBT1,HBT2}. They referred to the method as intensity interferometry. 
In particle physics, an analogous effect was discovered 
independently~\cite{GGLP} by G. Goldhaber, S. Goldhaber, W. - Y. Lee and 
A. Pais (GGLP), who observed an enhancement of identical pions
with small opening angles in $p+\overline{p}$ reactions. 
Essentially, the enhancement
of identical bosons with small momentum difference is due to their
Bose-Einstein (BE) statistics. The effect is frequently referred to as either 
the HBT, or the GGLP effect, or intensity interferometry, or simply as
 BE correlations (BEC).

The detailed shape analysis of the BEC is rather complicated
but important, because the shape carries information about the 
space-time structure of the particle emission process. Recently,
two of us proposed~\cite{edge} a model-independent method for the 
analysis of the shape of this function, based on an experimentally
preferred weight function and a complete orthogonal set of polynomials,
where orthogonality is defined with respect to a 
preferred weight function.
In the case of approximately Gaussian correlation functions,
the method results in the Edgeworth expansions, and the complete
orthogonal set of functions are the Hermite polynomials. 
In the case of an approximately exponential shape, 
the expansion is given in terms of the Laguerre polynomials. 
For approximately spherical distributions 
the expansion can be formulated in terms
of spherical harmonics 
(see ref.~\cite{edge} for further details).
 
The Edgeworth and the Laguerre expansions are very general
methods to characterize the two-particle correlation functions.
They depend only on the following two 

\underline{\it experimental conditions:}

{\it i) } 
{\it The correlation function tends to a constant
for large values of the relative momentum $Q$.}

{\it ii)}
{\it The correlation function deviates from its asymptotic, large $Q$ value
in  a certain domain of  its argument. }

For simplicity, this domain, where the correlation function
deviates from its asymptotic value, with other words,
the  location of its non-trivial structure,
is assumed to be close to $Q = 0$.

This implies that the Edgeworth, Laguerre (and similar) expansions 
can be applied to all kinds of data
that satisfy properties {\it i)} and {\it ii)}, as the bosonic
properties of the observed particles are not utilized.
Such expansions are useful if one would like to study and characterize 
short range correlations in a general manner, and if no attempt
is made to connect the structure of the shape of the BEC 
to the characteristics of the underlying space-time
picture.  (The conditions for the convergence of such expansions are given 
in ref.~\cite{edge}.)

The purpose of the present study is to provide a model-independent
characterization of the BEC with a further additional assumption:

{\it iii)} {\it The two-particle correlation function
 is related to a Fourier transformed
space-time distribution of the source.}

This last assumption implies that the plane-wave approximation
can be warranted, and additional possible short-range correlations
due to e.g. the final state Coulomb or strong interactions as well as
energy-momentum conservation or the jet structure of the source
can be removed from the data or their magnitude is under theoretical
and/or experimental control.

The purpose of the present paper is to determine in a general form
the approximate behaviour of the two-particle BEC 
under assumptions {\it i) - iii)}
and relate the parameters of the source functions to the parameters
of the BEC in a generic form.

\subsection{Basics}
The two-particle BEC is defined as the
ratio of the two-particle invariant momentum distribution
$N_2(\bk_1,\bk_2)$ to the 
product of the single-particle invariant momentum distributions
$N_1(\bk_i)$~\cite{misko},
where $\bk$ stands for the three-momentum of the detected particle
of mass $m$ and energy $E = \sqrt{m^2 + \bk^2}$.
The BECF is defined as 
\be
C_2(\bk_1,\bk_2) = \frac{N_2(\bk_1,\bk_2)}{N_1(\bk_1)\, N_1(\bk_2)},
	\label{e:cdef}
\ee
for more details on the normalizations see refs.~\cite{cs-rev,misko}.
For simplicity, we assume that at large relative momenta,
the two-particle correlation function tends to 1.
Thus we neglect possible long-range correlations
and normalization coefficients in eq.~(\ref{e:cdef}).

As $C_2(\bk_1,\bk_2)$ is a ratio of invariant momentum distributions,
the correlation function is also an invariant observable. This property
can be emphasized by using the $C_2(k_1,k_2)$ notation, where 
$k_1$ and $k_2$ are the momentum four-vectors of the particle pair.
See also refs.~\cite{cs-rev,kittel-rev} for more detailed 
discussions about the constraints between $k_1$ and $k_2$.

We assume that $C_2(k_1,k_2)$ is 
experimentally determined and  satisfies
properties {\it i) } - {\it iii)}.
These correlation functions 
can be evaluated at a certain fixed mean momentum
${\mathbf K} = 0.5 (\bk_1 + \bk_2)$ as a function of various
components of the relative momentum ${\mathbf q} = \bk_1 - \bk_2$. 
The relative momentum can be decomposed into some
experimentally preferred relative momentum components 
in 1, 2 or 3 dimensions. For example, 
experimental data are presented 
in one dimension in terms of the invariant momentum
difference $Q_I = \sqrt{ - (k_1 - k_2)^2}$,
or as a function of various components of the relative momentum,
for example, $(q_0, q_z,  q_t) $ = ($E_1 - E_2$, $\bk_{1,z} - \bk_{2,z}$, 
$\sqrt{ (\bk_{1,x} - \bk_{2,x})^2+ (\bk_{1,y} - \bk_{2,y})^2}$ ). 

Various directional decompositions of the relative momentum
with respect to the mean momentum have been proposed.
These decompositions include the Bertsch-Pratt, the Yano-Koonin-Podgoretskii
and the invariant Buda-Lund decompositions. For further details, we
refer the interested readers to the recent reviews on these topics,
\cite{cs-rev,kittel-rev,uli-b-rev,uli-rev,weiner-rev}.
For our further considerations, the specific experimental choice for
the directional decomposition of the relative momenta will not be
essential.

Essential background information on the generic structure of the
peak of the BEC was summarized in ref.~\cite{qinv-m}
regarding the impossibility of the Taylor expansion of the 
correlation functions around the $Q=0$ point, and 
in the appendix of ref.~\cite{3d}, where it was pointed out that
the class of the stable distributions plays an important role 
in non-Gaussian  BEC studies.
The break-down of the Gaussian approximation around the
$Q=0$ point  was shown with the help of
a theoretical analysis and a numerical simulation in ref.~\cite{dkiang}.
Explicit simulations, including long tailed and/or
asymmetric source density distributions 
confirmed this result~\cite{voloshin-halo}.

The above results are essentially the consequence of the fact that
the two-particle phase-space volume grows as $Q^3$ for small values
of the relative momentum $Q$. Hence the number of pion pairs tends to zero
as $Q \rightarrow 0$, which implies that  the numerator and the
denominator of eq.~(\ref{e:cdef}) tend to 0 in the same limit,
$\lim_{Q \rightarrow 0} C_2(Q) = "0/0"$. This is the reason why
it is experimentally impossible to determine $C(Q=0)$ {\it exactly}:
the experimental errors on the correlation function
diverge as $Q\rightarrow 0$.
(Not to mention $\frac{dC(Q)}{dQ}$ and $\frac{d^2C(Q)}{dQ^2}$
whose experimental errors diverge even more than that of $C(Q)$).
So the two-particle correlation function 
can only be measured at non-zero values
of $Q > Q_{min}$, at a  scale set by the combination of factors such
as the available statistics, the two-track resolution and the extent to which
final-state interactions are under control. Typical values of $Q_{min}$ are 
5-10 MeV in present correlation measurements in heavy ion physics.  
The value of the Bose-Einstein correlation function at $Q=0$ can then be estimated with various extrapolation techniques, e.g. with the help of 
a (typically Gaussian) model assumption for the 
functional form of $C(Q)$, or with the help of
model-independent Edgeworth or Laguerre expansions~\cite{edge}. 

\section{Bose-Einstein correlations for univariate stable distributions}

For clarity, let us consider first the
case of a one-dimensional static boson source, and show the
essential ideas in the simplest possible manner.
(In the subsequent sections we generalize these
results to more realistic, expanding sources that
contain correlations between coordinate space and momentum space, 
and also to the case of more complicated derivations.)

Thus we assume here that the source can be characterized by
a factorized form of the emission function:
\be
	S(x,k) =  f(x) \, g(k),
\ee
where $x$ and $k$ are one-dimensional coordinate and momentum variables,
respectively. The normalization is chosen such that
\be
	\int \d x \, f(x) \, = \, 1, \qquad\qquad
	\int \d k \, g(k)  =  \langle n \rangle, 
\ee 
where $\langle n\rangle$ stands for the mean multiplicity.
The single-particle spectrum is obtained as
\be
	N_1(k) = \int \d x \, S(x,k) = g(k).
\ee
Assuming chaotic particle emission, and the validity of the plane-wave
approximation, the two-particle Bose-Einstein symmetrized wave-function
reads as
\be
	\psi_{k_1,k_2}(x_1,x_2)  =  \frac{1}{\sqrt{2}}
	\left[\exp(i k_1 x_1 + i k_2 x_2) + 
		\exp(i k_1 x_2 + i k_2 x_1)\right].
\ee
In the Yano-Koonin formalism~\cite{yanokoonin}, 
the two-particle momentum distribution
is determined as
\be
	N_2(k_1,k_2) = \int \d x_1 \d x_2 \, S(x_1,k_1) \, S(x_2,k_2) \,
		|\psi_{k_1,k_2}(x_1,x_2)|^2,
\ee
and the resulting two-particle BEC is found to be
\be
	C_2(k_1,k_2) = 1 + |\tilde f(q_{12})|^2,
\ee
which measures the absolute value squared Fourier transformed
coordinate-space distibution function of the particle emitting source,
where the Fourier transformed source density (often referred to as the
{\it characteristic function}) and the relative momentum are defined as
\be
	\tilde f(q_{12})  =  \int \d x \, \exp(i q_{12} x) \,f(x),\qquad\quad \label{e:char}
	q_{12}  =  k_1 - k_2. \label{e:fourier}
\ee
As the two-particle Bose-Einstein 
correlation function is determined by the modulus of this Fourier-transformed
source distribution, but is insensitive to the phase of the Fourier transform,
these modulus squared Fourier transforms measure only 
the {\it relative coordinate} distributions of the source,
but they are not sensitive to shift parameters like the location of the center
of the source distribution in space-time.
It is frequently assumed (either explicitly or implicitly), 
that the function $\tilde f(q)$ is analytic
around $q=0$ and that its second order
Taylor expansion characterizes its behaviour well even for large
values of $q$.
 We shall see later that this imposes severe constraints
on the possible choice of the source functions. As this point is essential,
we emphasize that the analyticity of $\tilde f(q)$ at $q=0$ is 
{\it an assumption} that greatly reduces the possible forms of the 
source densities. This may be
clearly seen in the case of Levy stable
distributions\cite{zol2,zol1,nolan-chap1}, 
which are characterized by four real parameters,
and thus form an uncountably infinite set. The most
essential of these four parameters is the index of stability $\alpha$, which
may take any value in the interval $(0,2]$. However, the
assumption that the $f(q)$ characteristic function is analytic restricts this
parameter to the special  $\alpha = 2$  case, corresponding to 
Gaussian Bose-Einstein correlation functions.

{\bf \it If} the function $\tilde f(q)$ is an analytic function
at $q = 0$ and {\it if} the lhs of eq. ~(\ref{e:fourier}) can be
expanded into a convergent series, one obtains
\be
	\tilde f(q) \approx 1 + i q \langle x \rangle - q^2 \langle x^2\rangle/2 + ... \, ,
\ee
where
\be
	\langle x^n \rangle = \int \d x \, x^n f(x) .
\ee
	This, in turn, implies that
\be
	C(q) = 1 + |\tilde f(q) |^2 \approx 
	2 - q^2 (\langle x^2\rangle - \langle x \rangle^2)
	\approx 1 + \exp( - q^2 R^2),
\ee
	where a Gaussian source radius parameter $R$ is defined as
	the width of the source emission function,
\be
	R = \sqrt{\langle x^2\rangle - \langle x \rangle^2}.
\ee
	Such type of over-simplified derivations, repeated in
	multi-dimensional forms, are frequently utilized to ``prove"
	that the Bose-Einstein correlation function  ``has to have
	a Gaussian form", without investigating the {\it domain of 
	applicability of the approximations} that yield this result.

	In fact, a Gaussian ansatz for the two-particle
	BEC frequently gives a good first approximation
	to the data, in particular,
	if the statitistical precision of the date is not adequate
	to perform a detailed shape analysis. One of the reasons
	for such a behaviour could be that the emission of
	elementary particles is a rather complicated, stochastic process,
	in particular in collisions with very high energies. 
	If one assumes that there are many independent processes, 
	that shift by $\delta x$ the coordinate $x$, and that the the 
	final production point 
	is a sum of many, similarly
	distributed, random shifts,
	$x = \sum_i \delta x_i$, and the shifts are characterized
	by finite variances, then for such random variables
	the probability distribution for $x$ tends to a Gaussian one,
	if all the conditions of the Central Limit Theorem are satisfied.
	As the Fourier transform of a Gaussian is also Gaussian
	in such cases the expected shape of the two-particle BEC is
	also a Gaussian. 

	However, there are many random processes for
	additive random variables, where a limiting distribution
	exists, but the Gaussian version of the central limit theorem
	is not applicable. Such processes are characterized by large
	fluctuations, power-law like tails and the non-analytic behaviour
	of the characteristic function of the probability distribution
	for small values of its arguments. In mathematical statistics and
	probability theory, such distributions are frequently referred
	to as L\'evy, or stable distributions. 

	Let us approach the problem of the shape of the
	two-particle BEC without a reference to a particular model,
	from the more general point of view of mathematical statistics 
	and the (generalized) Central Limit Theorems. Concrete,
	model specific applications shall be presented 
	elsewhere~\cite{be-alphas}.

\section{L\'evy stable distributions and generalized central limit theorems}

	In physics, as well as in the theory of probability, 
	the probability distribution of a sum of a large number
	 of random variables is one of the important problems,
	as such distributions are frequently realized in Nature. 
	Limit distributions characterize the probability distributions
	of random processes in the limiting case when number of
	the elementary independent random subprocesses 	tends to infinity.

	Various forms of the Central Limit Theorem state, 
	that under some conditions, the sum of a large number of
	random variables behaves as a Gaussian distribution. Essentially,
	the conditions for the validity of the various central limit theorems
	are that the elementary distributions should have a finite mean
	and variance. The sum of a large number of such variables
	follows a Gaussian distribution, which in turn  is a 
	(frequently encountered) special case of limit distributions.
	
	For example, rapidity is a kinematic variable
	invariant under longitudinal Lorentz transformations. The final
	(observed) rapidity of a particle is a consequence of many
	random and independent actions, which can be considered as
	some rapidity shifts. In case of rescattering on a hadronic
	or on a partonic medium, the final rapidity of the particles
	is obviously a sum of many random rapidity shifts. 
	If these subprocesses have finite means and variances,
	then the central limit theorem implies that the observable 
	rapidity distribution has to be a Gaussian, regardless of
	the details of the probability distribution of rapidity
	shifts in the elementary collisions.

	The stable distributions are precisely those limit distributions
	that can occur in Generalized Central Limit theorems. Their study
	was begun by the mathematician Paul L\'evy in the 1920's. 
	A recent book by Zolotarev and Uchaikin~\cite{zol2} 
	contains over 200 pages of applications of stable distributions 
	in probabilistic models, correlated systems and fractals, 
	anomalous diffusion and chaos,
	physics, radiophysics, astrophysics, stochastic algorithms,
	financial applications, biology and geology. Stable distributions
	provide solutions to certain ordinary and fractional differential
	equations, and the breadht of their applications suggests  
	that they can be considered as a class of special 
	functions~\cite{zol2,zol1,nolan-chap1}.

\subsection{Characteristic functions}
	The Fourier transformed density distribution, eq.~(\ref{e:char}) is usually called 
	the characteristic function in mathematical statistics.
	The stable distributions are frequently given in terms of
	their characteristic functions, as the explicit formulas 
	of the  corresponding source density distributions 
	are known only in some special cases.

	The characteristic functions of univariate stable distributions 
	can be defined using various convention schemes. In this work,
	we utilize the notation of Nolan~\cite{nolan-chap1}. In this work, the
	characteristic function of univariate stable distributions
	is given in three different conventions, out of which we utilize
	the convention $S(\alpha,\beta,\gamma,\delta;1)$ as this leads
	to natural generalization of data parameterizations utilized before
	in Bose-Einstein correlations in high energy physics.
	In this convention, the characteristic function $\tilde f(q)$, with
	other words, the Fourier-transform of a probability density given by
	the $S(\alpha,\beta,\gamma,\delta;1)$ law, has 
	the following general form:
\bea
	\tilde f(q) & = & \exp\left(- \gamma^\alpha |q|^\alpha 
			+ i \beta \gamma^\alpha |q|^\alpha 
			\mbox{\rm sign}(q) \tan(\frac{\alpha \pi}{2}) 
			+ i q \delta \right),
			\quad \alpha \ne 1.
\eea
	Thus univariate stable distributions are characterized by four parameters:
	the L\'evy index of stability (or characteristic exponent) $0 < \alpha \le 2$,
	the skewness parameter $-1 \le \beta \le 1$, the scale parameter $0 < \gamma$
	and the location parameter $- \infty  < \delta < \infty $.
	We recommend to consult the book of J. Nolan for greater details and the
	$\alpha = 1$ special case,~\cite{nolan-chap1}, see in particular Fig. 1.3
	of this work for an illustration of the stable densities in the
	$S(\alpha,\beta,\gamma,\delta;1)$  parameterization~\cite{nolan-chap1}.

	Before discussing the univariate case in general,
	let us start with one of the simplest cases. 
	Mathematically, this corresponds to the selection of the 
	symmetric stable distributions. Their relationship
	with the Gaussian case is also the most apparent. 
	Let us also introduce physical units.
	Using the notation of Nolan~\cite{nolan-chap1}, 
	this choice corresponds to the $S(\alpha,\beta=0,\gamma = 
	R/2^{\frac{1}{\alpha}},\delta = x_0;1)$ 
	convention. 

	For symmetric stable distributions, 
	the Fourier transformed density distribution has a simple form,
\be
	\tilde f(q)=\exp\left( i q \delta -|\gamma q|^\alpha\right), 
		\label{e:fqs}
\ee
	where $|z|$ stands for the modulus (absolute value) of the variable $z$,
	and the support of the density function $f(x)$ is $(-\infty,\infty)$.
	More complicated cases are discussed in the subsequent parts.
	In order to simplify the results, and to present results that
	are similar to the ones used in data fitting in high energy
	physics, it is useful to re-scale the scale parameter
	$\gamma$ of the L\'evy distributions 
	and introduce a physical notation  as follows:
\be
 	R \, = \, 2^{\frac{1}{\alpha}}\,\gamma  
	\qquad\quad
	x_0  = \delta \qquad (\mbox{\rm if}\quad \beta = 0).\label{e:resc}, 
\ee 
	In the chosen $S(\alpha,\beta,\gamma,\delta;1)$
	notational system, for symmetric $(\beta = 0)$ stable distributions, 
	the parameter $\delta$ coincides with $x_0$, 
	the location parameter of the distribution.

	A generic property of the stable distributions is that their
	characteristic function is non-analytic at $q=0$ for all 
	values of the index of stability $\alpha <2$. For symmetric
	distributions, the following small $q$ expansion reflects
	this non-analytic behaviour:
\be
	\tilde f(q)\approx 1 +  i q x_0 -\frac{1}{2}|q R|^\alpha
\ee
	The analyticity of this expansion is restored only if $\alpha = 2$,
	corresponding to the Gaussian limit.

	L\'evy-stable distributions  yield the following,
	relatively simple form of the two-particle BEC:
\be
	C(q;\alpha) = 1 + \exp\left(-|q R|^\alpha\right).
		\label{e:c2stable}
\ee
	This  form that has an additonal parameter, 
	the index of stability $\alpha$,
	as compared to the usual Gaussian (or exponential) distribution, 
	where the value of $\alpha$ is fixed to 2 (or 1).

	Note that eq.~(\ref{e:fqs}) is not an approximation for the
	shape of the characteristic function around the $q=0$ point,
	but an exact and deep mathematical result, a consequence of the
	Generalized Central Limit Theorem, which gives the possible
	forms of the Fourier transform of the density distributions
	of the stable laws. Hence eq.~(\ref{e:fqs}) is valid for all
	values of $q$, and the corresponding form, eq.~(\ref{e:c2stable})
	for the two-particle BEC is well suited for a detailed shape 
	analysis in the physically resolvable large $q$ region.

	The simplicity of eq.~(\ref{e:c2stable}) is the reason why we
	utilized the  $S(\alpha,\beta,\gamma,\delta=x_0;1)$ convention
	with a rescaling of the $\gamma$ parameter to $R$:
	In cases of $\alpha=1 $  and $\alpha=2$ we obtain
	formulas that already had a number of applications
	in particle interferometry  studies in
	high energy physics~\cite{cs-rev,kittel-rev,weiner-rev}.

	The physical meaning of the index of stability is 
	that the symmetric stable distributions
	decrease for large values of the coordinate $x$ as a power-law,
\be
	f(x) \propto x^{-1-\alpha}, 
	\quad \mbox{\rm for} \quad x \gg R, \quad 0 <\alpha < 2,
\ee
	while in the $\alpha = 2$ special case corresponds to 
	Gaussian source distributions.

	In general, stable distributions with $ 0 <\alpha < 2 $ are long tailed 
	and are frequently related to the self-similarity of the generating mechanism. 
	For example, in QCD, the theory of strong interactions, 
	jets emerge within jets within jets etc, 
	a process characterized by a self-similar branching and 
	a fractal structure, where the fractal dimension of the source 
	is related to the anomalous dimension of 
	QCD~\cite{gustafson,dokshitzer-dremin,dewolf-dremin-kittel}.

	In the Lund model of hadronization,  a self-similar
	string fragmentation process determines the distribution
	of the directly produced particles~\cite{lund}.
	Among these, there is a big range
	of the distribution of the decay times of various resonances,
	and their decay may lead also to a power-law tail
	fluctuation of the effective source sizes~\cite{bialas}.
	       Numerically, similar results
        were obtained by Utyuzh, Wilk and W{\l}odarczyk in ref.~\cite{wilk}
        when considering Bose-Einstein correlations for sources with
        a fractal, power-law structure in space-time.

	A second order phase transition is expected 
	if a heavy ion collision passes through the critical
	point of the quark-gluon plasma - hadronic matter phase transition,
	throught the end-point of the phase-boundary along which the transition
	is of first order. In the vincinity of the critical point the 
	second order phase transition is expected to be signalled
	by power-law type of fluctuations in coordinate space.
	Coordinate distributions with power-law tails may obey the 
	Generalized Central Limit Theorem, which essentially means
	that these density distributions, in the limit of
	large number of independent sources,
	tend to a L\'evy stable distribution with 
	index of stability $\alpha < 2$.

\subsubsection{Some examples} 
\begin{description}
\item{1)}
{\it The Gaussian  distribution} corresponds to the case of $\alpha = 2$. 
	The  scale parameter is $R = \sqrt{\langle x^2\rangle - 
	\langle x \rangle^2}$, the standard deviation.  
	The density distribution and the corresponding BEC are 
\bea
	f(x) & = &\frac{1}{(2 \pi R^2)^{1/2} }
		\exp\left[-\frac{(x-x_0)^2}{2 R^2}\right], 
			\quad -\infty < x <\infty,\\
	C(q) 	& = & 1 + \exp\left( - q^2\, R^2 \right).
\eea
\item{2)} 
	{\it The Lorentzian (or Cauchy) distribution} corresponds to the 
	case of $\alpha = 1$. The  scale parameter $R$ is 
	the standard scale parameter of the Lorentz distribution.
	The density distribution and the corresponding 
 	BEC are	
\bea
	f(x) & = &\frac{1}{\pi } 
		\frac{R}{R^2 + (x-x_0)^2},
			\quad -\infty < x <\infty,\\
	C(q) 	& = & 1 + \exp\left( - |q \, R| \right).
\eea
	Such a form was frequently applied to describe
	the $Q_I$ dependence of two-particle BEC 
	in high energy heavy ion and 
	particle physics~\cite{weiner-rev,cs-rev}.
\item{3)}
	{\it The asymmetric L\'evy  distribution} is 
	a slightly more complicated, but still relatively simple case, 
	corresponding to $\alpha = 1/2,\beta=1$. 
	The density has the following one-sided distribution
\bea
	f(x) & = &\sqrt{\frac{R}{8\pi }} 
		\frac{1}{(x-x_0)^{3/2}}
		\exp\left(-\frac{R}{8(x-x_0)}\right),
			\quad x_0 < x <\infty,\\
	C(q) 	& = & 1 + \exp\left( - \sqrt{|q \, R|} \right).
\eea
	This density distribution has been considered,
	although in a different formalism, for the characterization 
	of two-particle BEC in ref.~\cite{qinv-m}.
\end{description}

\subsection{Asymmetric L\'evy distributions}

	Adopting Nolan's $S(\alpha,\beta,\gamma,\delta;1)$
	convention for the univariate L\'evy stable densities with the
	rescaling of the scale parameter $\gamma$ to $R$, eq.~(\ref{e:resc}) ,
	the characteristic function $\tilde f(q)$ has 
	the following general form:
\bea
	\tilde f(q) & = & \exp\left(- \frac{1}{2} R^\alpha |q|^\alpha 
			+ i \frac{\beta}{2} 
			R^\alpha |q|^\alpha 
			\mbox{\rm sign}(q) \tan(\frac{\alpha \pi}{2}) 
			+ i q x_0 \right),
			\quad \alpha \ne 1,
\eea
	and we recommend to consult the book of J. Nolan for the
	$\alpha = 1$ special case,~\cite{nolan-chap1}.

	This form of the Generalized Central Limit Theorem also
	implies that two-particle BEC-s, 
	measuring the modulus square of the Fourier transformed
	source distribution, are insensitive to the asymmetry parameter
	$\beta$ of the L\'evy stable distributions, as well as to the
	center $x_0$ of the source. 
	This is the direct consequence of the fact, that the two-particle
	Bose-Einstein correlations measure only the relative coordinate
	distributions, hence asymmetry or the position of the center
	of the distribution does not enter the observable
        correlations.

	If the source is characterized by a L\'evy stable distribution,
	the two-particle BEC can always be
	written in the form of eq.~(\ref{e:c2stable}) 
	{\it in the whole range of $q$}. 

	Note that for $\beta \ne 0$, the interpretation of the
	$\gamma$ and $\delta$ parameters of the 
	$S(\alpha,\beta,\gamma,\delta;1)$ is not straightforward,
	and this parameterization is not a continuous function of 
	the parameter $\alpha$ at the $\alpha=1$ point. 
	See ref.~\cite{nolan-chap1} for a more 
	detailed discussion on these points.

\subsection{Three-particle Bose-Einstein correlations 
	for stable distributions}

	Can one extract, at least  in principle, 
	the asymmetry parameter $\beta$ of the stable laws,
	using the methods of particle interferometry?
	In order to answer this question, let us consider the three-particle
	Bose-Einstein correlation function. In general, it may have a 
	complicated form~\cite{pcnhalo,cs-rev}. 
	However, if the particle emission is completely chaotic and if
	the plane-wave approximation can be warranted,
	then the three-particle BECF reads as follows:
\bea
	C_3(1,2,3) \,  = \, 1 + |\tilde f(1,2)|^2 + |\tilde f(2,3)|^2
	 & +  &|\tilde f(3,1)|^2 +  \nonumber \\
 	 & +  & 	2 {\mathcal R} 	
			\tilde f(1,2) \tilde f(2,3) \tilde f(3,1).
\eea
	where the symbolic notation $\tilde f(i,j) \equiv \tilde f(k_i-k_j)
	 \equiv \tilde f(q_{ij})$ has been introduced to simplify the equation.  

	The three-particle {\it cumulant} correlation function corresponds
	to the last term,
\be
	\kappa_3(1,2,3) = 2 {\mathcal R} \tilde f(1,2) \tilde f(2,3) 
		\tilde f(3,1),
\ee
	where the two-particle {\it cumulant} correlation function is
	defined as
\be
	\kappa_2(1,2) = |\tilde f(1,2)|^2 .
\ee
	Hence, the normalized and symmetrized ratio
\be
	w(1,2,3) = \frac{\kappa_3(1,2,3)}
		{2 \sqrt{\kappa_2(1,2) \kappa_2(2,3) \kappa_2(3,1)}}
\ee
	turns out to be a simple function of the asymmetry parameter
	of the L\'evy stable distributions:
\bea
	w(1,2,3) & =  &\cos\left\{\frac{\beta}{2} 
			R^\alpha \tan(\frac{\alpha \pi}{2}) 
			[\sum_{(i,j)}
			|q_{ij}|^\alpha 
			\mbox{\rm sign}(q_{ij})]  \right\}
			\quad \mbox{\rm for}\quad  \alpha \ne 1, \label{e:cosfi}
\eea
	(for the special case of $\alpha=1$, see again ref.~\cite{nolan-chap1}.)
	In the above equation, the summation is taken over the cyclic
	permutations, $(i,j) = (1,2)$, $(2,3)$, or $(3,1)$. 
	Note that the displacement parameter $\delta = x_0$ cancels
	from this result, too. However, this parameter, related to the
	center of the particle emitting source, is not considered as
	an essential quantity in the physics of particle 
	interferometry. For simplicity
	we may assume that the origin of our coordinate system is fixed by
	the condition $x_0 = 0$.

	Let us emphasize that the result of eq.~(\ref{e:cosfi}) is
	valid only within the plane-wave approximation and neglecting
	possible partial coherence in the source~\cite{pcnhalo,cs-rev}. 
	If these conditions hold, the asymmetry 
	parameter $-1 \le \beta \le 1$ can be determined
	from the relative momentum dependence of the normalized three-particle
	cumulant correlation function $w$,
	which is denoted also~\cite{uli-3pi,henning-3pi} as 
\be 
	w = \cos(\phi).
\ee
	The angle $\phi$ is directly proportional to $\beta$, 
	the asymmetry parameter of the stable
	distributions, and the coefficents of proportionality
	can be determined from the two-particle correlation data.
	Thus we have proven, that for a one-dimensional L\'evy
	stable distribution the asymmetry parameter of the source
	can in principle be determined from a combined use of the
	two- and three-particle BEC-s. 
	Thus all the essential parameters of the source density 
	can be reconstructed within this class of density distributions.

	We would like to note, that the cases of $\beta= +1$ and $\beta = -1$
	are particularly interesting, as in these cases the support of
	$f(x)$ is one-sided, corresponding to $[\delta,\infty)$ and 
	$(-\infty,\delta]$, respectively. Such kinds of L\'evy stable
	distribitions may have applications e.g. in the  description of the
	temporal processes in the collisions of particles at high energies:
	particle production cannot start before the initial collision 
	- hence the corresponding time interval has to be limited from below.

	After its parameters $\alpha,\beta$ and $\gamma$ are determined,
	a L\'evy stable density distribution $f(x)$ can 
	be straightforwardly evaluated using public domain numerical 
	packages~\cite{nolan-num}. 
	In most of the cases these functions are also known in closed form.
	The difficulty is that they can be expressed in terms of very
	general kind of functions, Fox's H-functions, see Schneider's
	papers for greater details~\cite{schneider-p,schneider-m}.

	In this sub-section we neglected a possible core-halo correction, 
	because it is straightforward to shown that 
	it cancels from the final result on $w$, eq.~(\ref{e:cosfi}).

	Note, that in the more general case of expanding sources,
	Bose-Einstein correlations are sensitive only to the so called
	lengths of homogeneity, which correspond to the local
	density distribution for particles that have a given momentum $k$.
	In order to reconstruct the complete picture of the density
	distribution for all particles, one has to perform a combined
	analysis of the single particle spectra and the two-particle
	BEC-s~\cite{3d,qm95,cs-rev}.

	For such type of expanding sources, the radius parameters
	are known to decrease with increasing (transverse) mass
	of the particles~\cite{cs-rev,kittel-rev}. This implies, that
	the sensitivity of three-particle correlations for the shape 
	parameters of the source distributions is enhanced when
	the heavier kaons are utilized instead of the pions, as follows
	if one evaluates eq. (\ref{e:cosfi}) using a reduced radius parameter.

\section{Multivariate symmetric L\'evy distributions}
	
	Multivariate L\'evy stable distributions have properties
	similar to the uni-variate case, their study is 
	even presently a research topic in the theory of mathematical
	statistics, see ref.~\cite{nolan-summ}.
	 In this section, we assume that the
	source is symmetric, to simplify the presentation.

	Let us denote the coordinate 3-vector by ${\bf x}$, and
	the time by the variable $t$, while a space-time
	four-vector is $x = (t,{\bf x})$ , a momentum space four-vector is
	$k = (E,{\bf k})$. 

	The simplest case corresponds to a static (non-expanding) source,
	that emits all the particles instantaneously. This is described
	by a factorized form of  the emission function,
\be
	S(x,k) =	f({\bf x})  g({\bf k}) \delta(t-t_0) .
\ee
	{\it If} the density distribution $f({\bf x}) $ is a multivariate
	stable distribution that features ellipsoidal contour lines, 
	than the reconstruction
	of the source density distribution becomes a one-dimensional,
	solvable problem~\cite{nolan-fitting}. 

	The corresponding three-dimensional, symmetric, sub-Gaussian
 	L\'evy-stable characteristic function is
\be
	\tilde f({\bf q}) =\exp\left( i {\bf q}{\bf x_0} - \frac{1}{2} 
		| \sum_{i,j=1}^3 R^2_{i,j} q_i q_j |^\frac{\alpha}{2} \right)
\ee
	where ${\bf x}_0$ stands for the center of particle emission,
	and $R^2_{i,j}$ corresponds to a symmetric matrix $R_{i,j}=R_{j,i}$
	with dimensions of radius squared; $\alpha$ is the index of stability
	similarly to the one-dimensional case, and ${\bf q} = 
	{\bf k}_1 - \bk_2$ stands for the relative momentum.

	Again, we have applied a rescaling as compared to the 
	$S(\alpha,0,\gamma,0;1)$ class of sub-Gaussian distributions,
	that are discussed in ref.~\cite{nolan-fitting}.
	The rescaling corresponds to the application of eq.~(\ref{e:resc})
	interpreting that equation in a matrix notation.
	 We chose a notation that is
	adopted to the convention used in correlation studies
	in high energy physics, so that the 
	Gaussian limiting case, $\alpha=2$, corresponds to the 
\be
	C({\bf q}) = 1 + \exp\left(- \sum_{i,j} R^2_{i,j} q_i q_j  \right) 
\ee
	conventional multivariate Gaussian forms. 

	Hence, for symmetric, sub-Gaussian multivariate L\'evy distributions,
	the two-particle BEC has the
	following simple form
\be
	C({\bf q}) = 1 + \exp\left(-
		| \sum_{i,j=1,3} R^2_{i,j} q_i q_j |^\frac{\alpha}{2} \right)
\ee
	The components of the  squared radius matrix are symmetric,
	$R^2_{ij}=R^2_{ji}$. The components of this matrix and  that of the
	relative momentum can be determined in the experimentalists'
	favourite reference frame. 
	The point is that $\sum_{i,j} q_i R^2_{ij} q_j$
	is a (non-relativistic) bilinear form, which is (non-relativistically)
	invariant for the choice of the coordinate system, 
	hence ${\bf q R}^2{\bf q} $ can be decomposed in any frame.	

	The ellipsoids are characterized by the inverse of the 
	$R^2$ matrix, which is denoted symbolically by $R^{-2}$,
	and frequently called the covariance matrix: 
\be
	\sum_j R^2_{ij} R^{-2}_{jk} = \delta_{ik}.
\ee
	The large distance behaviour of the density distribution
	can be analytically determined as follows:
\be
	f({\bf x}) \approx 
		(\sum_{ij} R^{-2}_{ij} x_i x_j)^{-\frac{1+\alpha}{2}},
\ee
	so the source density decreases as a power-law for sufficently
	large values of the arguments, with $0 < \alpha <2$. (The case
	of $\alpha = 2$ corresponds to Gaussian source distributions).

	The elliptically contoured density distributions
	can be calculated by integrating a one-dimensional 
	distribution~\cite{nolan-fitting}.
	The ellipsoidal contour lines can be transformed to
	spherical contours by realizing that the
	two-particle BECF is characterized by $R^2$, a  symmetric 
	matrix  $(R^2)_{ij} = (R^2)_{ji}$ , which always can be
	written as $R^2 = R\cdot R^T$. The matrix $R$ can be 
	considered as a sort of `square root' of the squared
	radius matrix. Physically, the matrix $ R$ gives the 
	orientation of the prinicipal axis of the ellipsoids of
	the density profile in coordinate space, and characterizes
	also the stretchings along the principal axis that
	are needed to transform a unit sphere to an ellipsoid
	corresponding to the contour lines. Hence the
	 bilinear form in the exponent of the BECF  can also be
	written as ${\bf q R}^2{\bf q} = |{\bf R q}|^2$.

	The density distribution is expressed,
	in $d > 1$ dimensions, for $\alpha < 2$, by eqs.
	~(1) and (6) of ref.~\cite{zolotarev-1d},
\bea
	f_\alpha(s({\bf x})) & = & \frac{1}{(2 \pi \det R^2)^{d/2} } 
		\int_0^\infty \d t \, t^{d-1} \, (t s({\bf x}))^{1-d/2} \, 
		 J_{d/2 -1}(t s({\bf x})) e^{-t^\alpha}, \label{e:j0} \\
	s({\bf x}) & = & | R^{-1} {\bf x} |
\eea
	where $J_i(z)$ is the $i^{\mbox{\rm th}}$ order Bessel 
	function of the first kind, the matrix $R^{-1}$ is 
	the inverse of the radius matrix $R$,
	and the function $s({\bf x})$ corresponds to 
	a scaling variable that is constant over
	the ellipsoidal surfaces characterizing the
	constant source densities.
	It is relatively easy to see the physical
	meaning of this equation in $d=3$ dimensions, 
	in the system of ellipsoidal expansions,
	where the radius matrix corresponds to	
	$R = \mbox{\rm diag}(X,Y,Z)$, its inverse is
	$R^{-1} = \mbox{\rm diag}(1/X,1/Y, 1/Z)$ and the 
	scaling variable of the ellipsoidally symmetric 
	source distribution is given by
\be
	s({\bf x}) \,  = \, | R^{-1} {\bf x} | \,  = \,   
	\sqrt{\frac{r_x^2}{X^2} + \frac{r_y^2}{Y^2} + \frac{r_z^2}{Z^2}}.
\ee
	Note that  such scaling variables arise naturally in
	ellipsoidally symmetric solutions of fireball hydrodynamics,
	that may correspond to non-central collisions of heavy ions
	in the non-relativistic energy domain~\cite{ellsol1,ellsol2,csell}.

	In addition to the above analytic result, eq.~(\ref{e:j0}),
	the case of multivariate symmetric Levy laws is discussed 
	in great details  in refs.~\cite{zolotarev-1d}
	and~\cite{nolan-anal}. [We don't recapitulate here the well-known
	results for multivariate Fourier-transforms of  Gaussian densities,
	corresponding to the case of $\alpha=2$.]
	From the point of view of the applications, it is important that 
	public domain software packages are available~\cite{nolan-num}, 
	that can be utilized to plot the density distribution of stable laws
	as a function of the scaling variable $s$, if the 
	parameter $\alpha$ is determined from a multi-dimensional
	fit to the two-particle BEC-s.

	Multi-dimensional BEC-s are frequently
	studied in high energy particle and heavy ion physics.
	For the sake of completeness, and the applicability of our results
	to these cases, we give the two-particle BECF in a general form,
	using the Wigner-function formalism of 
	refs.~\cite{pratt-cstjz,uli-rev,cs-rev}.

\subsection{Wigner-function formalism }
	From now on, $x$ and $k$ stand for four-vectors. Using this notation,
	most of the results for the univariate case can be generalized
	to the relativistic case. 

	In the Wigner-function formalism, the particle emission is characterized
	by the source or emission function $S(x,k)$.
 	In the non-relativistic limit, this function corresponds to 
	the time-derivative of the Wigner function, which itself is 
	the quantum mechanical analogy to the classical phase-space 
	distribution and this $S(x,k)$ can also be considered
	as a relativistic, covariant form of the Wigner function.

	The invariant momentum distribution and the 
	two-particle BEC can be expressed
	in the Wigner-function formalism as follows:
\bea
	N_1(k_1) & = & \int \d^4 x \, S(x,k_1),\\
	C_2(k_1,k_2) & = & 1 + \frac{|\tilde S(q,K)|^2}
			{\tilde S(0,k_1) \tilde S(0,k_2)}
	\approx 1 + \frac{|\tilde S(q,K)|^2}{|\tilde S(0,K)|^2},\\
	\tilde S(q, K) & = & \int \d^4 x \, \exp(i q x) S(x,K),\\
	K & = & \half (k_1 + k_2),\qquad \quad q \, = \, k_1 - k_2. 
\eea
	By introducing the notation
\be
	f_{K}(x) = \frac{S(x,K)}{\int \d^4 \, x S(x,K)},
\ee
	the results of the earlier section are recovered, 
	but with all the parameters of the effective source 
	depending on the mean momentum ${K}$.

\subsubsection{Core-halo correction}

	If some of the particles are emitted from the decays of 
	very long lived resonances, such that their decay width 
	is much smaller than the typical 5-10 MeV
	width of the first bin of the experimentally 
	determined two-particle BECF, then the decay products 
	of these particles contribute to a halo that
	cannot be resolved with the help of two-particle BEC 
	measurements~\cite{chalo}. 
	Within the core-halo picture, their effects can
	be taken into account by assuming that the single-particle
	emission function can be written as a sum of 
	a resolvable core contribution
	and an unresolvable halo contribution,
\be
	S(x, k)  = S_c(x, k) + S_h(x, k)
\ee
 	and  as a result, an intercept parameter $\lambda$
	is introduced to the two-particle BEC,
\be
	C_2(k_1,k_2)  \approx   
	1 + \lambda \frac{|\tilde S_c(q,K)|^2}{|\tilde S_c(0,K)|^2},
\ee
	where the {\it effective} intercept parameter 
	$0 \le \lambda \equiv \lambda(\bk) \le 1$ 
	is interpreted as a (mean-momentum dependent) 
	measure of the fraction of bosons emitted 
	directly from the core:
\be
	\lambda = \left[\frac{N_c(K)}
		{N_c({K}) + N_h({ K})}\right]^2,
\ee
	and the emission function of the core determines the
	relative-momentum dependence of the BECF.

	This formalism is relevant in case of correlation studies for
	expanding systems. Assuming that the local regions of homogeneity
	can be characterized by a symmetric multivariate L\'evy stable
	distribution, the general form for the two-particle BECF is as follows:
\be
	C_2(k_1,k_2) = 1 + \lambda 
		\exp\left[ - (\sum_{i,j=1}^3 
		R_{ij}^2 q_i q_j)^{\alpha/2}\right],
\ee
	where all the parameters may depend on the mean momentum of the pair,
	$(\lambda,R_{ij}^2,\alpha) = 
		(\lambda({\bk}),R_{ij}^2({\bf K}), \alpha({\bf K}))$.

	It is at this point where the {\it large $q$} behaviour of the 
	Fourier transformed core density distribution becomes essential:
	within the core-halo picture, the exact intercept parameter
	of the two-particle BECF is $\lambda_{\rm xct} = C_2(k,k)-1 = 1$,
	however, it is assumed that the halo of long lived resonances
	creates an unresolvable peak in the $q < q_{\rm min}$ region,
	where the two-particle relative momentum resolution is typically
	$q_{\rm min} \sim 5$~MeV/c. This peak makes the analytic
	expansions around $q=0$ impossible, and the large $q$ behaviour
	of the core correlator becomes important. Fortunately,
	this is uniquely determined with the help of the index of
	stability $\alpha$ for the chosen class of stable distributions.

	The resulting BEC-s 
	all have a generalized exponential form, but, as compared to the
	usual Gaussian case, there exist an additional scale transformation
	$q_i \rightarrow q_i^{\alpha/2}$ which sometimes is also referred to
	as stretching. Hence, the resulting correlation functions 
	are multivariate {\it stretched exponential} parameterizations. 
	For $\alpha=2$, the earlier, Gaussian forms are recovered, 
	while for $\alpha < 2$ the correlation function becomes
	more peaked than a Gaussian, and it develops longer tails as well. 
	As such a peaked, but almost Gaussian
	behaviour is frequently seen in the experimentally determined
	Bose-Einstein correlation data in high energy particle and heavy
	ion induced reactions, we may hope that the above stretched
	exponential or sub-Gaussian
	parameterization provide a useful tool for the characterization of
	such datasets even in the case of the multi-dimensional analysis.

	It follows from the above discussion, that perhaps the presentation
	of the two-particle Bose-Einstein correlation data in the usual,
	linear-linear plot of $C(q)$ versus $q$ is not particularly useful.
	In order to exhibit the possible power-law structure of the exponent
	of the correlator, a log-log plot of the two-particle cumulant
	correlation function, $\ln\left\{\ln[C(q)-1] - \ln\lambda\right\}$ 
	versus $\ln q$ would be the most appropriate. On such plots,
	the two-particle correlation functions would appear as straight lines
	and the slope parameter of the lines would be given by the 
	index of stability $-\alpha$.

	Below let us give two examples, where we highlight 
	the essential model assumptions and show the result. 
	For more detailed discussion on these topics,  
	see refs.~\cite{cs-rev,kittel-rev}.

\subsection{L\'evy stable  distributions for non-relativistic, 
	expanding sources}

	Let us consider  a popular, but not boost-invariant 
	parameterization of the two-particle BEC-s, 
	the  Bertsch-Pratt~\cite{bertsch,pratt}  
	or side, out, long decomposition,  
	where ${\bf q} = (q_{\rm s},q_{\rm o}, q_{\rm l})$,
	where $q_l = k_{1,z} - k_{2,z}$ is the longitudinal, $q_{\rm o} = 
	(q_x K_x + q_y K_y)/\sqrt{K_x^2 + K_y^2}$ is the out(ward) and
	$q_s = (q_x K_y - q_y K_x)/\sqrt{K_x^2 + K_y^2}$ is the side(ward)
	relative momentum komponent. This decomposition is frequently
	studied in the LCMS frame~\cite{lcms}, 
	where $K_z=0$, because in this frame, the temporal
	component of the source couples only to the out direction,
	see e.g.~\cite{cs-rev} for details.
	This can be shown by utilizing the mass-shell constraints for the
	four-momenta, from which it follows that in the LCMS
\be
	E_1 - E_2 = \frac{\bk_1 + \bk_2}{E_1 + E_2} ({\bk_1 - 	\bk_2})
	\, = \, \beta_o q_o .
\ee
        
	If correlations between coordinate and time are
	negligible, a model assumption may be that the emission function 
	at every value of the particle momenta can be
	written as a factorized coordinate-space distribution
	and a temporal distribution,
\be
	S(x,{\bf k}) = S({\bf x},{\bf k}) h(t)
\ee
	and we may assume that at every value of ${\bf k}$
	the coordinate space distribution corresponds to a 
	three-dimensional, sub-Gaussian L\'evy distribution with index of 
	stability $\alpha_x$, while the temporal distribution
	would correspond to a univariate L\'evy distribution
	with index of stability $\alpha_t$, and in general
	these two indexes may be different from each other.

	Such kind of factorization assumptions can be approximately
	made not only for non-relativistic expanding sources but
	sometimes also for relativistic expanding sources when
	the emission function is evaluated in the LCMS frame~\cite{lcms},
	and the motion is non-relativistic with respect to this frame.

	The resulting two-particle BECF for cylindrically
	symmetric sources can be written as 
\be
	C_2(\bk_1,\bk_2) = 1 + \lambda \exp\left[- 
		(\sum_{i,j=s,o,l} R_{ij}^2 q_i q_j)^\frac{\alpha_x}{2} -
		(\Delta t^2 \beta_o^2 q_o^2)^\frac{\alpha_t}{2}
		\right] \label{e:stable-bp}
\ee
	An important feature of eq.~(\ref{e:stable-bp}) is that the spatial
	and the temporal scales may enter the relative momentum dependence
	in the out direction with different powers, $\alpha_x$ and $\alpha_t$,
	hence, a detailed shape analysis of the BECF in
	terms of BP variables can be utilized to {\it extract} these different
	indexes as well as the information on both the spatial and the
	temporal scales. However, as is well known, the $\alpha_t = \alpha_x$
	case becomes degenerate, and only the combination
	$R_o^2 + \beta_o^2 \Delta t^2$ can be determined from the
	data analysis in this exactly Gaussian case.

	A more realistic case is to include possible 	
	correlations between production time and position,
	which goes beyond the above factorized ansatz.

\subsection{L\'evy stable  distributions for 
	the invariant Buda-Lund variables}
	
	For collisions with very high energy, the particle emission
	process becames a highly relativistic phenomena. In this case,
	the invariance of the emission function can be reflected
	if the longitudinally boost-invariant proper-time variable
	$\tau=\sqrt{t^2 - r_z^2}$ is utilized, and the space-time
	rapidity $\eta = 0.5 \log[(t + r_z)/(t-r_z) ] $ is also
	introduced as a hyperbolic, boost-additive coordinate.

	In a factorized form the Buda-Lund (BL) parameterization 
	assumes the following structure for the emission function~\cite{3d}:
\be
	S(x,k) = H_*(\tau) G_*(\eta) I_*(r_x,r_y),
\ee
	where the subscript $_*$ denotes an implicit momentum dependence.
	The effective proper-time and space-time rapidity distributions
	$H_*(\tau)$ and $G_*(\eta)$ are assumed to have a uni-variate 
	L\'evy distributions with indexes of stability
	$\alpha_=$ and $\alpha_{||}$, while $I_*(r_x,r_y)$ may have a bivariate
	L\'evy distribution with index $\alpha_\perp$. Here we refer to
	the symbolic notation of the invariant
	temporal and the parallel relative momentum differences,
	$Q_=$ and $Q_{||}$, being conjugated variables to 
	the space-time variables $(\tau,\eta)$.

	The corresponding factorized L\'evy-stable BL parameterization
	of the two-particle BECF reads as
\bea
	C(k_1,k_2) &=& 1 + \lambda \exp\left(
		- 
		|R_= Q_=|^{\alpha_=}
		- 
		|R_\parallel  Q_\parallel|^{\alpha_\parallel}
		- 
		|R_\perp Q_\perp |^{\alpha_\perp}
		\right), \\
	Q_= & = & 
		 m_{t,1} \cosh(y_1 - \overline{\eta}) 
			- m_{t,2} \cosh(y_2 - \overline{\eta})\\
	Q_\parallel & = & 
		 m_{t,1} \sinh(y_1 - \overline{\eta}) 
			- m_{t,2} \sinh(y_2 - \overline{\eta})\\
	Q_\perp & = & \sqrt{Q_x^2 + Q_y^2}.
\eea
	In these equations, $m_{t,i} = 
	\sqrt{m^2 + {\bf k}_i^2}$ is the transverse
	mass, $y_i = 0.5\ln[(E_i + k_{z,i})/(E_i-k_{z,i})] $ is the rapidity 
	of particle $i$ and $\overline{\eta}$ stands for the space-time
	rapidity of the point of maximum emissivity for particles with 
	a given fixed four-momentum $k_i$. Note that the above equations
	for the definition of the temporal and parallel invariant
	momentum differences are equivalent to the ones given in 
	ref.~\cite{cs-rev}, but the invariant decomposition is perhaps more
	apparent in the above form.

	The three different indexes of stability, 
	$\alpha_=$, $\alpha_{\parallel}$ and $\alpha_\perp$ 
	appear as new fit parameters, due to 
	the assumed factorization property of the boost-invariant
	temporal, longitudinal  and transverse distributions from
	each other. All are L\'evy indexes for the corresponding
	spatial distributions, hence satisfy the usual unequality
	$0 < \alpha_i \le 2$ for all $_i = ( \null_=, \null_{\parallel},
	\null_{\parallel})$.

	All the five scale parameters, $\lambda$, $R_=$,
	$R_{\parallel}$, $R_{\perp}$	and $\overline{\eta}$
	may depend on the value of the mean momentum $K$, due to the
	expanding nature of the particle emitting source.

\section{Applications to NA22 and UA1 data}

	The case of the uni-variate stable distributions can be
	checked, relatively easily, against the precision
	data on the two-particle short-range correlation functions
	as measured in the CERN experiments NA22 ~\cite{na22-d2s}
	and UA1~\cite{ua1-d2s}.

	We emphasize, that we are not attempting here to reconstruct
	the detailed space-time structure of the  particle emission
	process, or, interpret the result in terms of a physical
	process that may result a L\'evy stable law. 
	We merely consider this application as a  first test of 
	the feasibility of our method as a quantitative tool
	for the  analysis of two-particle BEC 
	in high energy physics. 

	The parameters for the best L\'evy stable fits
	are summarized in Table 1.

	The two-particle BEC 
	for L\'evy stable distributions is a stretched exponential.
	Hence the title of Figure 1 where the best L\'evy stable fits
	are shown for both the NA22 and the UA1 data sets.
	
	The plots of the best L\'evy fits 
	are clearly better than even the best exponential fits
	to the same curve (not mentioning the "best Gaussian" 
	approximations, who miss completely the strongly rising 
	peak near to the $Q=0$ point).

	The quality of the fits is approximately similar to those
	obtained with the help of Laguerre expansions.	
	However, an advantage of the L\'evy stable laws and
	fits is that they relate the exponent $\alpha$ to the 
	power-law tail behaviour of density fluctuations in the
	large coordinate-space distributions of elementary particles.
	We discuss a possible interpretation of our result in
	terms of QCD cascades in a separate paper~\cite{be-alphas}.

	The following statements can be made without any reference to
	a dynamical model:

	Both data sets can be described by the L\'evy stable fits
	in a statistically acceptable manner. 
	In case of the NA22 data, the confidence level of the fit
	is about 80 \% even when one of the fit parameters is fixed,
	as determined from a Laguerre expansion method in ref.~\cite{edge}.
	Without fixing this parameter, the minimum was found to be
	degenerate and the confidence level even higher.
	In case of the UA1 data, the confidence level of the fit is
	only 6 \%, a rather low value.

	In case of the NA22 data set, the fitted value of the 
	$\lambda $ parameter is within errors close to the maximum 
	value of $1$ which is a theoretical upper limit within 
	the core-halo picture. Such an upper limit is violated by the fit
	to the UA1 data set. This may indicate that in the UA1 data
	there some non-Bose-Einstein type of short range correlations
	present. However, fixing $\lambda=1$ a L\'evy stable fit
	to the UA1 data becomes still possible, but only  with a
	small a confidence level of about 1 \%. 
	Fixing the normalization to that of the
	Laguerre expansion for the same function, ref.~\cite{edge},
	a L\'evy stable fit becomes possible with only 
	a very small, 0.1 \% confidence level.
	In all of the above cases, $\alpha$ remains significantly below 1.

	Fixing $\alpha=1$ (exponential fit), the confidence level 
	of the fit to 	the NA22 data  decreases from 80\% to 24 \%,
	in case of the UA1 data, from 6 \% to 10$^{-8}$.
	Fixing $\alpha=2$ (Gaussian fit), the  confidence level of 
	the best fit to the NA22 data decreases to $10^{-8}$,
	while it becomes essentially 0 in case of the UA1 data set.

	The conclusion is that L\'evy stable (or stretched exponential)
	fits describe the NA22 and the UA1 data on short-range correlations
	of particles in high energy physics. In case of the NA22 data,	
	the confidence level of the fit is very high and the fitted parameters
	can be interpreted within the core-halo picture, with a
	fully resolved halo. In case of the UA1 data, the best fits
	are obtained when the parameters were utilized outside 
	of their usual  domain. This result may indicate
	the presence of short-range correlations of not Bose-Einstein type in
	the UA1 data set.

\section{Discussion and conclusion}

	L\'evy stable laws appear in many branches of physics, mathematics,
	biology, economy, computer science and other areas, where the
	scale of fluctuations may be characterized by long tails and
	an asymptotic power-law like behaviour. As the L\'evy stable
	distributions satisfy a generalized central limit theorem,
	they may appear naturally also in case of high energy elementary
	particle physics, where self-similar branching processes are
	known to exist on the large momentum transfer scale, a self-similar
	string fragmentation process is expected to describe the 
	distributions of directly produced hadrons, and where the 
	distribution of abundances of various resonances may also
	follow an approximate power-law like behaviour 	as a function of the
	life-time of the resonance. Similar processes are expected
	to exist also in high energy heavy ion collisions, where additional
	fluctuations from a hadronic or quark matter may also modify
	the spectrum of these fluctuations. 
	
	We have argued, that additive stochastic processes
	may determine the space-time density distribution of 
	elementary production in highly energetic reactions
	of particles or heavy ions. In this case, if a generalized
	Central Limit Theorem is applicable, the probability
	distribution of particle emission corresponds to 
	a(n univariate or multivariate) L\'evy stable distribution.

	Although the characterization of the L\'evy stable distributions
	is not straightforward, by now these distributions are numerically
	as well as analytically accessible. Their most important
	property is that the Fourier transformed emission function
	has in general a non-analytic structure in the vanishing
	relative momentum region. 

	We have shown, that due to this reason, the two-particle
	Bose-Einstein correlation functions are characterized by a
	sub-Gaussian two-particle Bose-Einstein correlation function,
	with an index of stability $0 < \alpha \le 2$. The analytic
	behaviour and the Gaussian limiting case corresponds to 
	the $\alpha=2$ special case only, a single point out of 
	the uncountably infinite possibilities in the parameter space.
	Hence there is no reason to believe that the two-particle
	Bose-Einstein correlation functions have to behave
	as (multivariate) Gaussian forms. 

	We have demonstrated, in the simplest one-dimensional case,
	that the NA22 and the UA1 two-particle Bose-Einstein
	correlation functions are well described with a
	L\'evy-stable law and the index of stability was
	found to be significantly smaller than 2, i.e. the
	deviation from a Gaussian behaviour in this case
	can be quantified and characterized with our method.

	We have also determined the form of the
	two-particle Bose-Einstein correlation function 
	in case the particle emission is characterized by a
	multivariate L\'evy-stable law. In particular, we have
	generalized the non-relativistic
	Bertsch-Pratt and the relativistically invariant
	 Buda-Lund parameterizations for multivariate 
	L\'evy-stable laws.

\section*{Acknowledgments}
Cs. T. is indebted to W. Kittel for inspiration, support and
careful attention. Without him this manuscript could not have been
formulated. He is also greatful to R. Hakobyan and the L3 collaboration
for an inspiration to complete this paper,  and to J. P. Nolan
for enlightening communications.
This research has been supported by an NWO - OTKA grant N25487,
by the Hungarian NSF grants OTKA (T024094, T026435), T034269, T038406 and
T043514, by a NATO Senior Fellowship (Cs. T.), the NATO PST.CLG.980086 grant
and by an MTA-OTKA-NSF grant.
\vfill\eject

\vfill\eject
\begin{table}[hbt]
\newlength{\digitwidth} \settowidth{\digitwidth}{\rm 0}
\catcode`?=\active \def?{\kern\digitwidth}
\caption{Best fits to UA1 and NA22 two-particle correlations 
using a L\'evy stable law}
\label{tab:results}
\begin{center}
\begin{tabular*}{\textwidth}{@{}|l@{\extracolsep{\fill}}|rl|rl|}
\hline
                 & \multicolumn{2}{l|}{UA1} 
                 & \multicolumn{2}{l|}{NA22} \\
\cline{2-5}  
		 Parameter~\,~\,~\,
                 & \multicolumn{1}{r}{Value} 
                 & \multicolumn{1}{l|}{Error} 
                 & \multicolumn{1}{r}{Value} 
                 & \multicolumn{1}{l|}{Error} 
		 \\
\hline
${\cal N}$      & 1.34	&$\pm$ 0.03 	& 0.95 \,\, (fixed) \\
$\lambda$ 	& 1.85  &$\pm$ 0.07 	& 1.15 &$\pm$ 0.17 \\
$R$ [fm]  	& 6.36  &$\pm$ 0.33	& 1.33 &$\pm$ 0.30 \\
$\alpha$  	& 0.49  &$\pm$ 0.01	& 0.67 &$\pm$ 0.07 \\
\hline
$\chi^2/NDF$     & \multicolumn{2}{l|}{56.5/42 = 1.35} 
                 & \multicolumn{2}{l|}{27.67/35 = 0.769} \\
CL 		 & \multicolumn{2}{l|}{6.6 \%}
                 & \multicolumn{2}{l|}{80.6\%} \\
\hline
\end{tabular*}
\end{center}
\end{table}

\vfill\eject
\begin{figure}[ht]
\centering
\epsfig{file=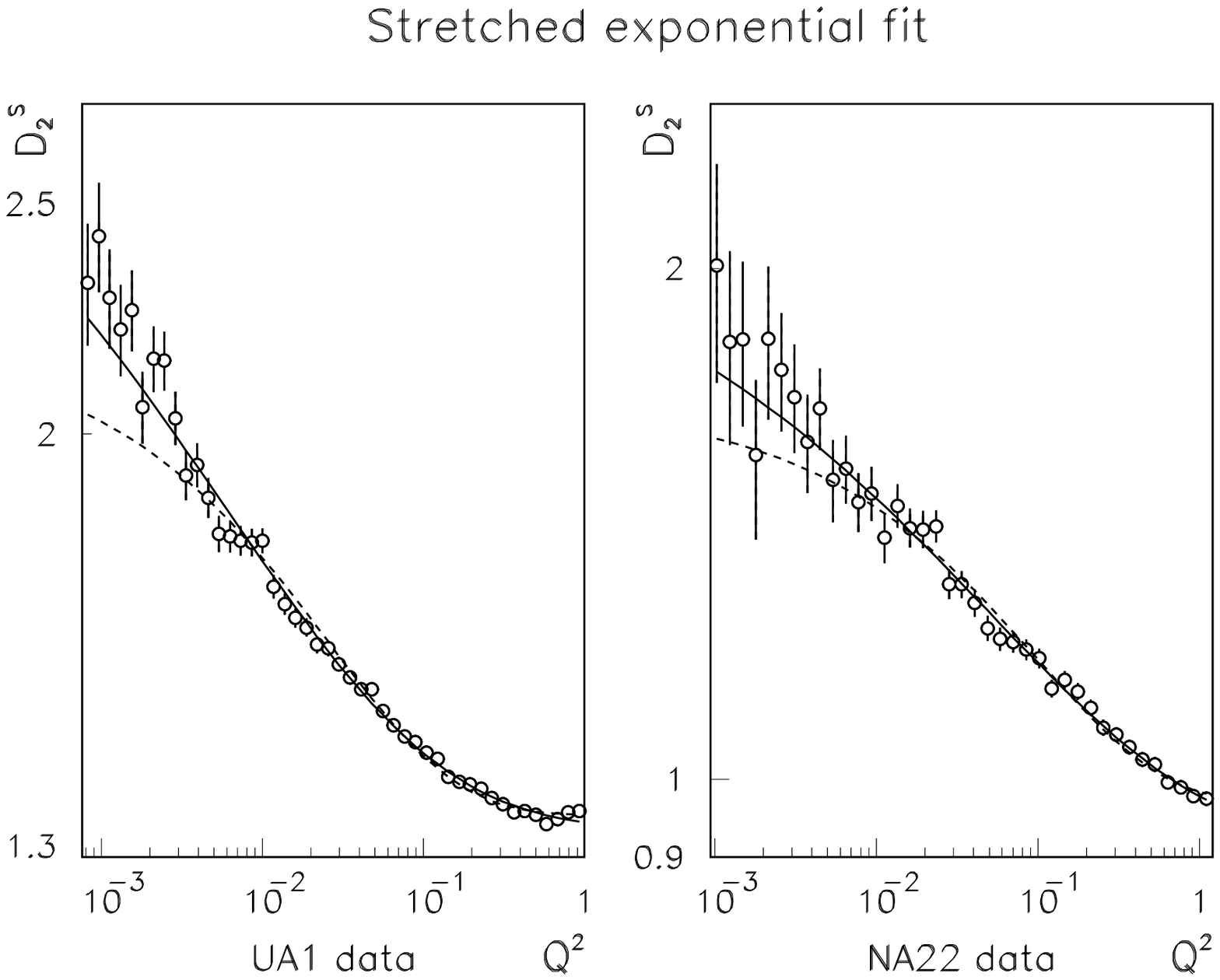,height=10cm}
\label{f:sexp}
\caption{The figures show $D_2^s$ which is proportional to
the two-particle short range correlation 
function, as measured by the UA1 and the NA22 collaborations at CERN.
The lines stand for the stretched exponential (or Levy) fits,
which are able to reproduce the data with an 
acceptable $\chi^2/NDF$, while the dashed lines
stand for the result of the best exponential fits.
The fit results are summarized in Table 1.
}
\end{figure}
\vfill\eject 

\begin{thebibliography}{99}
\bibitem{chalo}
T.~Cs\"org\H o, B.~L\"orstad and J.~Zim\'anyi,
Z.\ Phys.\ {\bf C71} (1996) 491
hep-ph/9411307.

\bibitem{bec-ua1}
S.E.~Vance, T.~Cs\"org\H o and D.~Kharzeev,
Phys.\ Rev.\ Lett.\  {\bf 81} (1998) 2205,
nucl-th/9802074.

\bibitem{hubble}
A. Ster and T. Cs\"org\H{o},
{\it  PRHEP}-hep2001 (2001) 241
[arXiv:hep-ph/0112065], 
see also \\
M.~Csan\'ad, T.~Cs\"org\H{o}, B.~L\"orstad and A.~Ster,
arXiv:nucl-th/0403074, J. Phys. G in press.

 

\bibitem{pcnhalo}
T.~Cs\"org\H o, B.~L\"orstad, J.~Schmid-Sorensen and A.~Ster,
Eur.\ Phys.\ J.\ {\bf C9} (1999) 275
hep-ph/9812422.

\bibitem{cs-rev} 
T.~Cs\"org\H{o},
Heavy Ion Phys. {\bf 15}  (2002) 1-80,
[arXiv:hep-ph/0001233] 

\bibitem{kittel-rev}
W. Kittel, Acta Phys.\ Polon.\ B {\bf 32} (2001) 3927,
[arXiv:hep-ph/0110088]



\bibitem{HBT1}
R.~Hanbury Brown and R.~Q.~Twiss,
Phil.\ Mag.\  {\bf 45}, 663 (1954).
 
\bibitem{HBT2} 
R.~Hanbury Brown and R.~Q.~Twiss,
Nature {\bf 178}, 1046 (1956).
 
\bibitem{GGLP} 
G.~Goldhaber, S.~Goldhaber, W.~Y.~Lee and A.~Pais,
Phys.\ Rev.\  {\bf 120}, 300 (1960).
 

\bibitem{edge}
T.~Cs\"org\H{o} and S.~Hegyi,
Phys.\ Lett.\ B {\bf 489}, 15 (2000).

\bibitem{misko}
D.~Miskowiec and S.~Voloshin,
Heavy Ion Phys.\  {\bf 9}, 283 (1999)
[arXiv:nucl-ex/9704006].
 

\bibitem{uli-b-rev}
U.~W.~Heinz and B.~V.~Jacak,
Ann.\ Rev.\ Nucl.\ Part.\ Sci.\  {\bf 49}, 529 (1999)
[arXiv:nucl-th/9902020].

\bibitem{uli-rev} 
U.~A.~Wiedemann and U.~W.~Heinz,
Phys.\ Rept.\  {\bf 319}, 145 (1999)
[arXiv:nucl-th/9901094].
 

\bibitem{weiner-rev} 
R.~M.~Weiner,
Phys.\ Rept.\  {\bf 327}, 249 (2000)
[arXiv:hep-ph/9904389].

\bibitem{qinv-m}
T.~Cs\"org\H{o} and J.~Zim\'anyi,
Nucl.\ Phys.\ A {\bf 517}, 588 (1990).
 
\bibitem{3d}
T.~Cs\"org\H{o} and B.~L\"orstad,
Phys.\ Rev.\ C {\bf 54}, 1390 (1996)
[arXiv:hep-ph/9509213].

\bibitem{dkiang}
S.~Nickerson, T.~Cs\"org\H o and D.~Kiang,
Phys.\ Rev.\ {\bf C57} (1998) 3251
[arXiv:nucl-th/9712059].

\bibitem{voloshin-halo}
D.~Hardtke and S.~A.~Voloshin,
Phys.\ Rev.\ C {\bf 61}, 024905 (2000)
[arXiv:nucl-th/9906033].

\bibitem{wilk-utyuzh}
	

\bibitem{yanokoonin}
	F. B. Yano and S. E. Koonin,
	Phys. Lett. B {\bf 78} (1978) 556 


\bibitem{zol2}
	V. V. Uchaikin and V. M. Zolotarev,
	{\it ``Chance and Stability, Stable Distributions and 
	Their Applications"}, VSP Science,1999, ISBN: 90-6764-301-7, 596 pp. 

\bibitem{zol1}
	V. M. Zolotarev, {\it ``One-dimensional Stable Distributions"},
	{\it  Am. Math. Soc. Transl. of Math. Monographs}, vol. {\bf 65},
	Providence, R.I. (Transl. of the original 1983 Russian)


\bibitem{nolan-chap1}
	J. P. Nolan, {\it Stable Distributions: Models for Heavy Tailed Data}
	\hfill\\
	{\tt http://academic2.american.edu/$\tilde{\,\,\,}$jpnolan/stable/CHAP1.PDF}	

\bibitem{be-alphas} 
	T. Cs\"org\H{o}, S. Hegyi and W. Kittel, manuscript in preparation.

\bibitem{nolan-fitting}
	J. P. Nolan, {\it Fitting Data and Assessing Goodness-of-fit
	with Stable Distributions}, Contribution ot the Heavy Tails
	Conference, \hfill \\
	{\tt http://academic2.american.edu/$\tilde{\,\,\,}$jpnolan/stable/DataAnalysis.ps}	

\bibitem{gustafson}
	G.~Gustafson and A.~Nilsson,
	Nucl.\ Phys.\ B {\bf 355} (1991) 106.

\bibitem{dokshitzer-dremin}
	Y.~L.~Dokshitzer and I.~M.~Dremin,
	Nucl.\ Phys.\ B {\bf 402} (1993) 139.

\bibitem{dewolf-dremin-kittel}
	E.~A.~De Wolf, I.~M.~Dremin and W.~Kittel,
	Phys.\ Rept.\  {\bf 270}, 1 (1996)
	[arXiv:hep-ph/9508325].

\bibitem{lund}
	B. Andersson, G. Gustafson, G. Ingelman, T. Sj\"ostrand,
	{\it Phys. Rept.} {\bf 97} (1983) 31
 
\bibitem{bialas}
	A.~Bialas,
	Acta Phys.\ Polon.\ B {\bf 23}, 561 (1992).

\bibitem{wilk}
        O.~V.~Utyuzh, G.~Wilk and Z.~Wlodarczyk,
        Phys.\ Rev.\ D {\bf 61} (2000) 034007
 

\bibitem{uli-3pi} 
U.~W.~Heinz and Q.~H.~Zhang,
Phys.\ Rev.\ C {\bf 56}, 426 (1997)
[arXiv:nucl-th/9701023].

\bibitem{henning-3pi}
	H.~Heiselberg and A.~P.~Vischer,
	arXiv:nucl-th/9707036.

\bibitem{zolotarev-1d}
	V. M. Zolotarev,
	in  {\it Contributions to probability}, Academic Press, New York, 1981,
	pp. 283-305

\bibitem{nolan-anal}
	H. Abdul-Hamid and J. P. Nolan, 
	{\it J. of Multivariate Analysis} {\bf 67} 80-89 (1998) ;\\
	J. P. Nolan and B. Rajput,
	{\it Commun. Statistics-Simula. } {\bf 24} 551-556 (1995).

\bibitem{nolan-num}
	J. P. Nolan, 
	a program to calculate multivariate stable densities;\hfill	\\
	{\tt http://academic2.american.edu/$\tilde{\,\,\,}$jpnolan/stable/mvstable.exe}\\

\bibitem{schneider-p}
	W.R. Schneider: 
	{\it Stable distributions: Fox function
	representation and generalization},
	in Stochastic Processes in Classical and Quantum Systems
	(Lecture Notes in Physics, Springer Verlag, Berlin, 1986,
	eds. S. Albeverio, G. Casati and D. Merlini), 
	Vol. {\bf 262}, pp. 497-511.
	

\bibitem{schneider-m}
	W.R. Schneider: 
	{\it Generalized one-sided stable distributions}
	Proceedings of the second BiBoS Symposium,
	in: Stochastic Processes - Mathematics and Physics II.
	(Lecture Notes in Mathematics, 
	Springer Verlag, Berlin, 1986,
	eds. S. Albeverio, Ph. Blanchard and L. Streit)
	Vol. {\bf 1250} pp. 269-287.  

\bibitem{qm95}
	T.~Cs\"org\H{o} and B.~L\"orstad,
	Nucl.\ Phys.\ A {\bf 590}, 465C (1995)
	[arXiv:hep-ph/9503494].

\bibitem{nolan-summ}
	J. P. Nolan, {\it Multivariate stable distributions: approximation,
	estimation, simulation and identification}, \hfill \\
	{\tt http://academic2.american.edu/$\tilde{\,\,\,}$jpnolan/stable/overview.ps}	


\bibitem{ellsol1}
	S.~V.~Akkelin, T.~Cs\"org\H{o}, B.~Luk\'acs, 
	Y.~M.~Sinyukov and M.~Weiner,
	Phys.\ Lett.\ B {\bf 505} (2001) 64
	[arXiv:hep-ph/0012127].

\bibitem{ellsol2}
	T.~Cs\"org\H{o}, S.~V.~Akkelin, Y.~Hama, B.~Luk\'acs and 
	Y.~M.~Sinyukov,
	arXiv:hep-ph/0108067.
 
\bibitem{csell} 
	T.~Cs\"org\H{o},
	arXiv:hep-ph/0111139.


\bibitem{pratt-cstjz} 
	S.~Pratt, T.~Cs\"org\H{o} and J.~Zim\'anyi,
	Phys.\ Rev.\ C {\bf 42}, 2646 (1990).
 
\bibitem{bertsch} 
G.~F.~Bertsch,
Nucl.\ Phys.\ A {\bf 498}, 173C (1989).
 
 
\bibitem{pratt}
S.~Pratt,
Phys.\ Rev.\ Lett.\  {\bf 53}, 1219 (1984).
 

\bibitem{lcms}
	T. Cs\"org\H o and S. Pratt, KFKI - 1991 - 28/A, p. 75

\bibitem{na22-d2s} 
N.~M.~Agababian {\it et al.}  [EHS/NA22 Collaboration],
Z.\ Phys.\ C {\bf 59}, 405 (1993).
 

 
\bibitem{ua1-d2s}
N.~Neumeister {\it et al.}  [UA1-Minimum Bias-Collaboration],
Z.\ Phys.\ C {\bf 60}, 633 (1993).
 

\end{thebibliography}
\end{document}